# A simple model for nanofiber formation by rotary jet-spinning

Paula Mellado[1,4], Holly A. McIlwee[1,2], Mohammad R. Badrossamay[1,2], Josue A. Goss[1,2], L. Mahadevan[1,3,4, a)], Kevin Kit Parker[1,2, a)]

[1]School of Engineering and Applied Sciences, Wyss Institute of Biologically Inspired Engineering, Harvard University, Cambridge, MA  02138
[2]Disease Biophysics Group, Harvard Stem Cell Institute, Harvard University, Cambridge, MA  02138
[3]Department of Physics, Harvard University, Cambridge, MA  02138
[4]Kavli Institute for NanoBio Science and Technology, Harvard University, Cambridge, MA 02138

**Nanofibers are microstructured materials that span a broad range of applications from tissue engineering scaffolds to polymer transistors. An efficient method of nanofiber production is Rotary Jet-Spinning (RJS), consisting of a perforated reservoir rotating at high speeds along its axis of symmetry, which propels a liquid, polymeric jet out of the reservoir orifice that stretches, dries and eventually solidifies to form nanoscale fibers. We report a minimal scaling framework complemented by a semi-analytic and numerical approach to characterize the regimes of nanofiber production, leading to a theoretical model for the fiber radius consistent with experimental observations. In addition to providing a mechanism for the formation of nanofibers, our study yields a phase diagram for the design of continuous nanofibers as a function of process parameters with implications for the morphological quality of fibers.**

The combination of high surface area ($10^3$ m$^2$/g),[1] mechanical flexibility, and directional strength make nanofibers an ideal platform for a diverse range of applications.[1-3] While nanofibers are most commonly produced using electrospinning,[4,5] we have recently demonstrated that Rotary Jet-Spinning (RJS) can be used as an alternative technique to fabricate sub-micron fibers using rotational inertial forces to extrude viscous polymer jets.[6,7,8] Our apparatus consists of a perforated reservoir containing polymer solutions attached to a motor (Fig. 1(a)). When the reservoir is spun about its axis of symmetry at a rate larger than a threshold determined by the balance between capillary and centrifugal forces, a viscous jet is ejected from a small orifice (Fig. 1(b)). This jet is thrown outwards along a spiral trajectory as solvent evaporates, owing to its relatively high surface area (Fig. 1(c-f)). While moving, it is extended by centrifugal forces (Fig. 1(g-j)) and solvent evaporates at a rate $J$, dependent on the diffusion coefficient $D$ of solvent through the polymer (Fig. 1(k)). The jet travels until it reaches the walls of the stationary cylindrical collector of radius $R_c$. Once there, the remaining solvent evaporates, fibers solidify, and may be collected.

To quantify the dynamics of jet spinning and understand the relative role of surface tension, inertial, non-inertial and viscous forces, we use scaling ideas to distinguish between three main stages of the process: (1) jet initiation, (2) jet extension, and (3) solvent evaporation. These three stages are characterized by the ejection time scale $\tau_1 \sim 1/\Omega_{th}$ proportional to the inverse of the critical angular speed $\Omega_{th}$ for which the jet is able to come out of the orifice (Fig. 1(c)), the viscous time scale $\tau_2 \sim r^2\rho/\mu$ of the polymer with extensional viscosity $\mu$ and density $\rho$ (Fig. 1(g)), and the solvent evaporation time scale $\tau_3 \sim r^2/D$ controlling the internal diffusion of solvent through the drying polymer (Fig. 1(k)). The ratio $\tau_2/\tau_3 \sim 10^{-2}$ marks a clear separation between the extension and solvent evaporation stages.

High speed imaging shows average time for a fiber to reach a collector of radius $R_c$ = 13.5 cm, when the reservoir is rotated at an angular speed $\Omega \sim 12,000$ rpm, $t_{gap} \sim 4\times10^{-2}$ s, which is consistent with the simple scaling $t_{gap} \sim 1/\Omega \sim 10^{-2}$ s. Measurements of solvent evaporation rate $J$ provide evidence that under ambient conditions, the solvent evaporates after a time $t >> t_{gap}$. This is because solvent evaporation is dominated by internal solvent diffusion through the polymer to the surface of the liquid jet[9] over a time scale $\tau_3 \sim r^2/D$, where $D \sim 10^{-7}$ cm$^2$/s is the diffusion coefficient of the solvent in the polymer[10] and $r$ is jet radius. From dimensional arguments it is evident that at a solvent concentration of 90 wt%, for instance, the ratio between the initial and final volume of the jet, when solvent has evaporated is $v_f = 0.1 v_{in} \sim 10$. Thus, in this approximation, the radius $r$ decreases to 1/3 the initial value if all the solvent evaporates. Fig. S2[14] shows that for $r$ to decrease to 1/3 its initial radius $r_0 \sim a/3$, the time elapsed is $t_{r0/3} \sim$ 250 s. Setting $\tau_3 = t_{r0/3}$, yields $r \sim a/3$ and consequently $t_{gap}/\tau_3 \sim 10^{-4}$. To demonstrate the timescale of solvent evaporation, fiber mass was measured as solvent evaporated after spinning. Evaporation was visualized by incorporation of a solvent sensitive dye into polymer solutions prior to spinning (Fig. 1(l)).[11,14] Both theory and experiment show that most solvent evaporates after fibers have landed on the collector walls. Therefore, jet extension dominates the process of controlling fiber radius.

The first stage of fiber formation is jet initiation. Since the hydrostatic pressure $\rho g h$ at the orifice is three orders of magnitude smaller than the centrifugal force $\rho \Omega^2 s_0 a$ at the orifice of radius $a$ in the reservoir of radius $s_0$, its role in jet ejection is negligible. Then, balancing inertial force $\rho \Omega^2 s_0 a^3$ with the capillary force at the orifice on the reservoir, $\sigma a$, yields a critical rotational speed for jet ejection $\Omega_{th} \sim \sqrt{\sigma/a^2 s_0 \rho}$, and an ejection speed $V \sim \Omega_{th} s_0$. Once the polymer jet exits the orifice, its extension is driven by the balance between viscous and centrifugal forces in the second stage of fiber formation. Noting that the viscous elongational stress on a jet of radius $r$ and velocity $V$ being stretched axially scales as $\mu V/x$, where $x$ is the distance from the orifice, and mass conservation implies that $Ua^2 = Vr^2$, we get an expression for the jet radius as a function of the experimental parameters that reads

$$r \sim \frac{aU^{1/2}\nu^{1/2}}{R_c^{3/2}\Omega}, \qquad (1)$$

where $a$ is the initial jet radius, $\nu = \mu/\rho$ is the kinematic viscosity, and $x \sim R_c$ is the collector radius ($R_c \gg s_0$).

To test Eq. (1), poly(lactic acid) fibers were fabricated using different experimental conditions.[14] The radius of the collector $R_c$, viscosity $\mu$, angular speed $\Omega$, and orifice radius $a$ were changed to span a range of fiber radii of one decade. PLA concentration varied from 4 to 9.5 wt% in chloroform, Ω from 4,000 to 37,000 rpm, and $R_c$ from 9 to 18 cm. Polymer concentration changed the viscosity of prepared samples from 25 to 250 mPa·s.[14] Surface tension and density of the samples remained approximately constant as polymer concentration was changed.[14] Fig. 2(a) shows fiber radius versus Eq. (1). This scaling law can be used to predict fiber radius and is shown to hold for multiple experimental schemes (Fig. 2(a)). The data are best fitted by the power law $\left(\dfrac{a^2 U \nu}{\Omega^2 R_c^3}\right)^{1/2} \sim r^{1.09 \pm 0.05}$ marked by the dashed line in Fig. 2(a), in good agreement with Eq. (1).

To obtain additional information regarding how the size of the jet scales with $\mu$ and $\Omega$, the dynamics of extension were examined in further detail by studying a simplified one-dimensional (1-D) theory for the fluid jet, assumed to be incompressible, Newtonian and isothermal. Making use of its slenderness,[12] we use a 1-D model set of equations in the axial coordinate $x$ for the conservation of mass and momentum[13,14] including effects of capillarity, inertial, viscous, and external body forces. Solving the resulting 1-D transient model numerically, we see there is good agreement (Fig. S1[14]) between experimental results, the simple steady state scaling law (Eq. S4[14]), and the transient solution.

The range of variability of $\mu$ and $\Omega$, which permit fiber formation, is summarized in the phase diagram in Fig. 2(b). Across the range of $\mu$ and $\Omega$ investigated, three regimes exist experimentally. In regime I, where $\Omega$, $\mu$ or both are large enough, continuous fibers, defined as those samples with less with 5% beads or defects, are produced. For example, at high angular speeds, $\Omega \sim 30,000$ rpm, continuous fibers are formed using concentrations between 4 wt% and 9 wt% spanning one decade in solution viscosity (regime I) (Fig. S2(a)[14]). Scanning electron microscopy (SEM) images show fiber morphology of samples within regime I: Fine continuous nanofibers collected from medium viscosity solutions (Fig. 2(d,g)), and large continuous microfibers collected from high viscosity solutions (Fig. 2(e,h)). Open data points in Fig 2(b) mark the lower limit of $\Omega$ for which continuous fibers prepared from solutions across the range of concentrations were collected.

As we decreased the viscosity of the solution, it was necessary to increase the angular velocity in order to produce collectable, continuous fibers. This was the case at 5.8 wt% and $\Omega$ = 8,000 rpm, where continuous fibers were produced (regime I). As $\Omega$, $\mu$, or both decrease to moderate values, regime II emerges where fibers can be fabricated, but beads or other defects are also produced. When the same 5.8 wt% solution was spun at $\Omega$ = 6,000 rpm, beaded fibers resulted (regime II). Beads or defects form because the centrifugal force is not large enough to overcome surface tension and elongate the jet before it reaches the collector. SEM images show beaded fibers collected in regime II from low viscosity solutions spun at high speeds (Fig. 2(c,f)). Filled data points in Fig. 2(b) mark the transition from regime II where beaded fibers are fabricated to regime III where $\Omega$, $\mu$, or both are too small to fabricate fibers. For all solutions in the range of polymer concentrations described above spun at $\Omega$ = 2,000 rpm, $a$ = 230 $\mu$m and $R_c$ = 13.5 cm, no fibers formed (regime III). At $\Omega$ = 3,000 rpm, fiber jets could be visualized leaving the reservoir though no solid fibers could be collected. In this case the centrifugal force was not enough to facilitate the jet reaching the collector (regime III). The range of $\Omega$ and $\mu$ explored in the phase diagram represents the parameter space necessary to produce both continuous and beaded fibers from solutions of PLA in chloroform based on rheological behavior.

The scaling law (Eq. 1) describes how fiber radius shrinks as we move toward larger $\Omega$ and $\mu$, inside regimes I and II. The transition between regime II and III can be explained with dimensional arguments in terms of the time for capillary induced breakup of a stationary jet $t_c \sim \frac{\mu r}{\sigma}$.[9] Comparing this time scale with that for the advective motion of the jet from the reservoir to the collector, $t_{gap} \sim 1/\Omega$, and using Eq. (1) for the jet radius, we find that the minimum angular speed for fiber formation is $\Omega_c \sim \frac{\rho R_c^2 \sigma^2}{a^2} \mu^{-3}$. This power law determines the transition between the regimes where no fibers can be created (III) and where fibers with beads or malformations are produced (II), consistent with experimental observations as shown in Fig. 2(b). As expected, the boundary between regimes II and III is best fit by the power law $\Omega \sim \mu^{-2.88 \pm 0.25}$ as shown by the continuous curve in Fig. 2(b).

Our combined experimental and theoretical study of the formation of polymeric fibers produced by RJS shows that fiber radius follows a simple scaling law (1) that characterizes how RJS-manufactured nanofiber radius is tuned by varying viscosity, angular speed, distance to the collector, and the radius of the orifice by studying forces governing the stages of fiber formation. A phase diagram for the design space of continuous nanofibers as a function of process parameters is in good agreement with experiments and has implications for the production rates as well as in the morphological quality of fibers with radius ranging from 150 nm – 3 microns.

**Acknowledgements** We acknowledge financial support of this work from Wyss Institute for Biologically Inspired Engineering, Harvard University Materials Research Science and Engineering Center (MRSEC), Harvard University Nanoscale Science and Engineering Center (NSEC), Defense Advanced Research Projects Agency (DARPA W911NF-10-1-0113), Harvard Center for Nanoscale Systems (CNS). H.A.M acknowledges the National Science Foundation Graduate Research Fellowship Program. P.M. and H.A.M. contributed equally to this work.

a) Authors to whom correspondence should be addressed. Electronic email: kkparker@seas.harvard.edu and lm@seas.harvard.edu

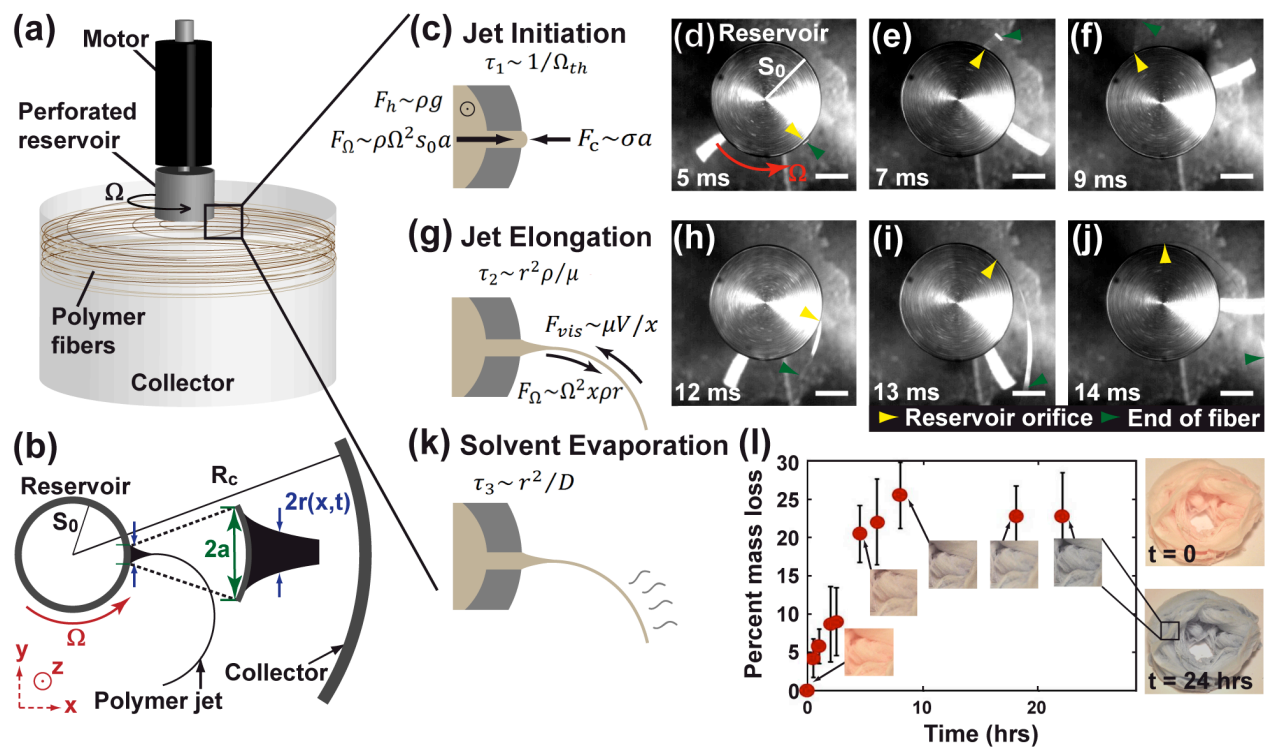

Fig. 1 (a) Schematic of RJS apparatus. (b) Top view diagram of a fiber projecting from the reservoir, towards the collector. (c) Jet initiation. (d-f) Photographic images capture initiation of the jet from the reservoir. Green arrows denote the end of the jet and yellow arrows mark orifice position. (g-j) Jet elongation. In (d-f,h-j) $s_0 = 0.85$ cm and scale bars are 0.42 cm. (k) Solvent evaporation. (l) Fiber color is an indication of the presence of solvent in the collected fibers, as solvent evaporates, fiber color changes from red to blue. Color evolution is in good agreement with the measured mass change of the collected fibers.

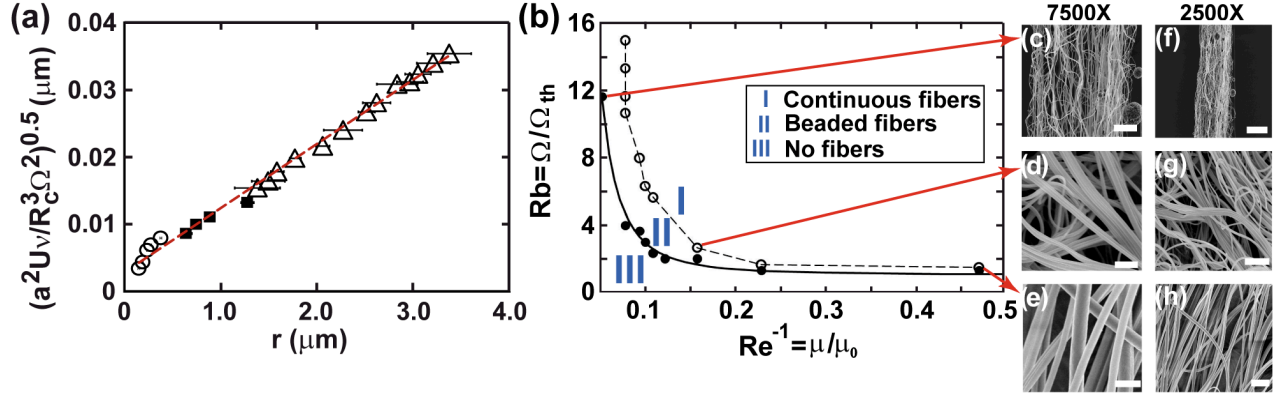

Fig. 2 (a) Fiber radius measurements. Circles: $a$ = 230 $\mu$m. Squares: $R_c$ = 13.5 cm, $\mu$ = 113 mPa·s, and $a$ = 230 $\mu$m. Triangles: $R_c$ = 13.5 cm, $\mu$ = 113 mPa·s, and $a$ = 410 $\mu$m. Data are best fitted by $(a^2Uv/R_c^3\Omega^2)^{0.5} \sim r^{1.09\pm0.05}$ (red dashed line). (b) A phase diagram divides the scaled angular velocity – viscosity (the Rossby number, $Rb = \Omega/\Omega_{th}$, $\Omega_{th}$ being the threshold speed for fibers to exit the reservoir, and the Reynolds number, $Re = \mu / \mu_0$, where $\mu_0 = \sqrt{\sigma s_0 \rho}$) plane into regimes I, II, and III. Open circles: transition to collecting beady fibers (regime II). Solid circles: transition to values at which no fibers can be collected (regime III). The boundary between regime II and III best fits the function $\Omega_c \sim \mu^{-2.88\pm0.25}$ (black curve). (c-h) SEM images showing fiber morphology of samples from different regimes at 2500X and 7500X magnifications. (c,f) Beady fibers are collected in II ($\mu$ = 0.025 Pa·s, $\Omega$ = 35,000 rpm, 8% beads). (d,g) Fine continuous fibers collected in I ($\mu$ = 0.079 Pa·s, $\Omega$ = 8,000 rpm). (e,h) Large continuous fibers collected in I ($\mu$ = 0.178 Pa·s, $\Omega$ = 4,800 rpm). Scale bar is 4 $\mu$m (c,d,e) and 20 $\mu$m (f,g,h).

**A simple model for nanofiber formation by rotary jet-spinning**

Paula Mellado[1,4], Holly A. McIlwee[1,2], Mohammad R. Badrossamay[1,2], Josue A. Goss[1,2], L. Mahadevan[1,3,4, a)], Kevin Kit Parker[1,2, a)]

[1]School of Engineering and Applied Sciences, Wyss Institute of Biologically Inspired Engineering,
[2]Disease Biophysics Group, Harvard Stem Cell Institute, Harvard University, Cambridge, MA 02138
[3]Department of Physics, Harvard University, Cambridge, MA 02138
[4]Kavli Institute for NanoBio Science and Technology, Harvard University, Cambridge, MA 02138

Discussion S1:

Description of experiments and apparatus. Experiments were performed using a Rotary Jet-Spinning (RJS) system as shown in Fig. 1(a) in the paper. The RJS system consisted of a brushless, servo motor (Maxon Motor Company, Fall River, MA) to which a custom designed perforated reservoir containing polymer solutions is attached (Fig. 1(a) in the paper). Rotation speed was held constant by a digital speed controller. The aluminum reservoir with an inner radius $s_0 = 0.85$ cm, and outer radius of 1.35 cm was used to contain the polymer solutions in all experiments. The reservoir had two sidewall orifices with radius $a = 230$ $\mu$m for all experiments, unless otherwise noted. Poly (lactic acid) (PLA, polymer 2002D, NatureWorks®, Minnetonka, MN) was dissolved in chloroform (99.8%, Mallinckrodt Chemicals, Phillipsburg, NJ) at concentrations ranging from 2 to 9.5 wt% at ambient conditions. Solution volumes of 2 ml (height inside reservoir was ~ 0.8 cm) were loaded into the perforated reservoir before rotation began. The motor was rotated at angular speeds $\Omega$ ranging from 0 to 37,000 rpm. Nanofibers collected on the stationary, round collector wall of radius $R_c$, were sputter coated with Pt/Pd using a Cressington 208HR sputter coater (Watford, England) to minimize charging during imaging using a Zeiss Supra field-emission scanning electron microscope (Carl Zeiss, Dresden,

Germany). Fiber diameter was analyzed using ImageJ image analysis software (National Institutes of Health, US).

Phase diagram. To construct the phase diagram of fiber quality, polymer concentrations ranging from 2 to 8 wt% were spun at Ω from 3000 to 52000 rpm. Samples were collected, sputter coated, and imaged as previously described. The phase diagram is composed of points marking the lower limit for which samples were able to be collected, and the transition from samples consisting of fibers and beads or other malformations of polymer to samples consisting of continuous fibers with less than 5% beads or malformations. Percentage beads or malformations were calculated using a custom manual thresholding macro using ImageJ. A total of *n* = 24 fields of view at 2500X and 7500X magnification were analyzed per condition.

Solvent evaporation. In order to visualize solvent evaporation, a solvent sensitive dye, 10,12-Pentacosadiynoic acid, was added to polymer solutions. In the presence of organic solvents (in this case chloroform) the blue dye turns red due to molecular backbone reorganization. Upon evaporation, the dye returns to its original blue color.[1] The progressive color change and solvent evaporation rate over 24 hrs is shown in Fig. 1(l) in the paper.

Discussion S2:

The one-dimensional problem was studied by considering a fixed coordinate system (*x, y, z*) having its origin at the center of the orifice as illustrated in Fig. 1(b). We considered an axial velocity gradient alone, meaning that the spinning solution experiences an elongational flow along the *x* direction only. Effects of air drag were neglected, and based on our measurements (see Evaporation Rate Note S4) we neglect solvent evaporation.

Equation of continuity along the *x* coordinate in steady state reads as follows:

$$\partial_x (VA) = 0, \tag{S1}$$

where *V* is the axial velocity and *A* is the jet cross sectional area. Solution of the continuity equation yields $Ur_0^2 = Vr^2$, where *U* is the initial axial velocity and $r_o$ and *r* define the initial and final radius of the jet respectively. Momentum balance along the *x* direction in the steady state reads:

$$\partial_x (V^2 A) = \partial_x \left( \frac{3\mu A}{\rho} \partial_x V + \frac{\sigma \sqrt{A}}{\rho} \right) + A\Omega^2 x, \tag{S2}$$

where *μ, ρ,* and *σ* define the fluid viscosity, density, and surface tension respectively. The momentum balance equation cannot easily be solved analytically, but we assume that the velocity field is smooth and monotonic when the fluid has just came out of the orifice. This allows us to find an algebraic expression for the radius of the jet in steady state as a function of the flow properties, experimental parameters and the axial coordinate *x*:

$$r/r_0 \approx \sqrt{\frac{\rho U x}{\mu - \Sigma + \sqrt{(\mu - \Sigma)^2 + (\rho \Omega x^2)^2}}}, \tag{S3}$$

where $r_0 \sim a$ is the radius of the jet at the orifice and $\Sigma = \frac{\sigma x}{r_0 U}$. In the previous formula it is easy to see that the radius size is determined by a competence between, viscous (proportional to *μ*), capillary (proportional to *σ*), and centripetal (terms proportional to Ω) forces.

Based on our scaling estimates, we expect the key variables controlling the jet diameter to be *μ* and Ω. Therefore we made use of equation (S3) and studied the ratio $r(R_c, \rho, \sigma, U, a, \kappa\Omega,$ $\kappa^2\mu)/r(R_c, \rho, \sigma, U, a, \Omega, \mu)$ where all parameters are kept constant except by $\mu$ and $\Omega$, which we scaled by the constants $\kappa^2$ and $\kappa$ respectively. We found that this ratio fit well with the function $\gamma/\kappa$, where $\gamma$ is a constant to be fixed depending on the other experimental parameters. In the limit of high angular speeds, $\Omega \rightarrow \infty$, $\gamma = 1/2$, yielding the following universal scaling relation:

$$r(\ldots,\kappa\Omega, \kappa^2\mu)/r(\ldots,\Omega, \mu) \sim 1/2\kappa. \tag{S4}$$

The right hand side of equation (S4) quantifies the decrease in jet radius when viscosity and angular speed are scaled by $\kappa^2$ and $\kappa$ respectively. We ran experiments to test equation (S4) where viscosity and rotation speed were scaled by the constants $\kappa^2$ and $\kappa$, respectively with $\kappa$ varying between 1 and 2.6, in order to test viscosity and angular speed values inside the experimentally viable range (thus polymer concentration varied from 5.45 wt% to 9.1 wt% and the angular speed in the range 15,000 to 36,000 rpm). Fig. S1 shows the experimental, analytical (steady state), and numerical (transient state) ratio $r(\ldots,\kappa\Omega, \kappa^2\mu)/r(\ldots,\Omega,\mu)$ as a function of the scaling factor $\kappa$. The analytical and numerical curves agree and are best fitted by the function $\gamma/\kappa$, with $\gamma \sim 0.6$ and are in fair agreement with the experiments.

Discussion S3:

The full equations of mass and momentum conservation along $x$ read as follows:

$$\partial_t A + \partial_x (VA) = -\frac{1}{\rho x}\frac{\delta M}{\delta t}, \tag{S5}$$

$$\partial_t(VA) + \partial_x(V^2 A) = \partial_x\left(\frac{3\mu A}{\rho}\partial_x V + \frac{\sigma\sqrt{A}}{\rho}\right) + A\Omega^2 x. \tag{S6}$$

Where $\frac{\delta M}{\delta t} = J$ is the solvent evaporation rate, $V=V(x,t)$ is the velocity field and $A=\pi\, r(x,t)^2$ is the jet cross sectional area. The first and second terms of the right hand side of equation (S6) are the one-dimensional stresses given by the viscous and surface tension. The boundary conditions at the reservoir orifice read:

$$V(0,t) = U; r(0,t) = a; \partial_x V(R_c, t) = 0, \tag{S7}$$

and the initial conditions:

$$r(x, 0) = a; V(x, 0) = U, \tag{S8}$$

where $U$ is the initial axial velocity of the jet at the orifice in the rotating coordinate system due to the hydrostatic pressure on the reservoir ($U \sim 10$ cm/s). We numerically solved the partial differential equations (S5) and (S6) using the software Mathematica 7.0 with the boundary and initial conditions provided by equation (S7) and equation (S8) respectively. We employed the Method of Lines, which consisted of discretizing the spatial dimension, and then numerically integrating the temporal variable as an ordinary differential equation. We made use of the "TensorProductGrid" Method for spatial discretization and a pseudospectral derivative approximation both available in Mathematica 7.0.

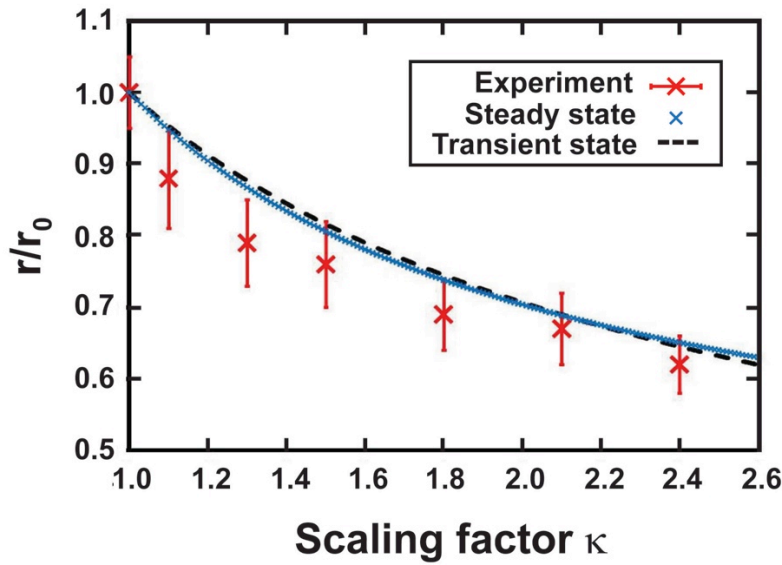

Fig. S1. Experimental (large red *x*), analytical (black dashed line), and numerical (small blue *x*) ratio $r(..., \kappa\Omega, \kappa^2\mu)/r(..., \Omega, \mu)$ as a function of the scaling factor $\kappa$. The theoretical curves are best fitted by the function $\gamma/\kappa$, with $\gamma \sim 0.6$ and are in good agreement with experimental data. Error bars represent the standard deviation of the sample mean. $R_c = 13.5$ cm, $\rho \sim 1.54$ g/cm$^3$, $\sigma \sim 27$ g/s$^2$, $U \sim 10$ cm/s and $a = 170$ $\mu$m in all experiments. Polymer solutions were composed of PLA dissolved in chloroform at varying viscosities $\mu$, as noted.

Discussion S4:

In order to quantify the change in solution viscosity as the jet travels to the collector, the rate at which chloroform is evaporated from a solution containing PLA was determined by recording the mass of solutions over time. Evaporation rate of polymer solutions ranging from 5.8 – 12 wt% contained in glass vessels ranging in diameter from 1.9 – 4.6 cm was measured. The rate of evaporation was found to be equal to $\frac{\delta M}{\delta t} = 0.15$ g/min, independent of the initial polymer concentration and container size. Consequently, the concentration of solvent, $c_{fs}$, after a time $t$ has elapsed changes according to $c_{fs} = (c_{is} - 0.0035t)/(1 - 0.0035t)$ for an initial solvent concentration $c_{is}$.

Discussion S5:

The viscosity $\mu$, as a function of polymer concentration $c_p$, was measured, Fig. S2(a). To determine the concentration regimes, rheological measurements were made on freshly prepared PLA solutions. Solutions were loaded into the viscometer (Model AR-G2, TA instruments, New Castle, DE) fitted with a standard-size recessed end concentric cylinders geometry (model 988339, stator outer radius 14 mm, rotor inner radius 15 mm, 4000 $\mu$m gap) and viscosities were measured under steady state shear rate from $0.1 - 3000$ s$^{-1}$.

The experimental viscosity versus polymer concentration curve is best fitted by a polynomial function for small concentrations of polymer and by an exponential function at higher concentrations. The best fit gives: $\mu = 0.68c_p^2 + 5.02c_p + 0.01\exp(c_p)$. These results were used to find the approximate behavior of $\mu$ with respect to initial solvent concentration and time, which are shown in Figs. S2(a) and S2(b). For illustrative purposes for an initial solvent concentration $c_{is} = 2$ wt%, we have computed the change in viscosity during $t_{gap}=10^{-2}$s to find that $\Delta\mu(t = t_{gap}) = \mu(0.92,0) - \mu(0.92, t_{gap}) \sim 0.06$ mPa·s which is a negligible 0.05 percent of the viscosity's value at the beginning as can be verified by Fig. S2(b).

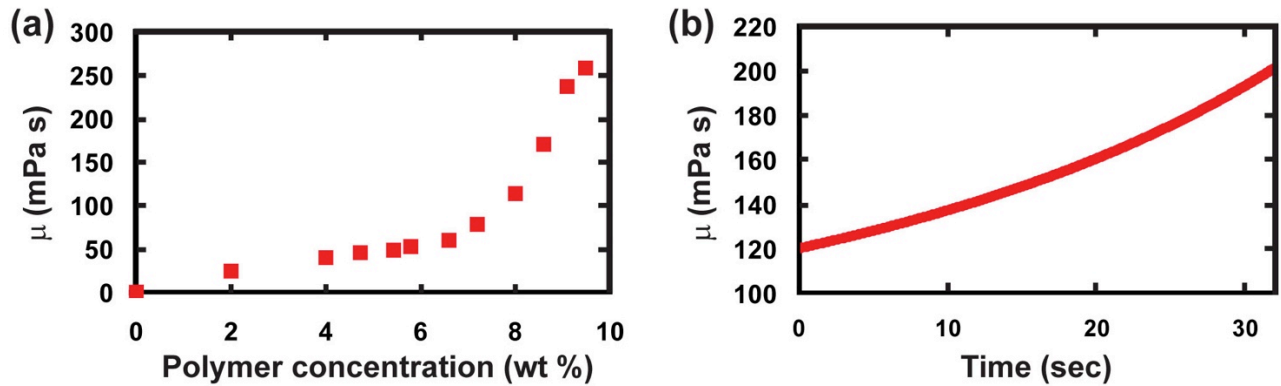

Fig. S2. Experimental measurements and theoretical approximations to determine viscosity of solution jets during the RJS process. (a) Experimentally measured shear viscosity of solutions of PLA dissolved in chloroform as a function of polymer concentration. (b) Approximated behavior of the viscosity of PLA solutions versus time, derived from the measured evaporation rate of chloroform from solutions and measured shear viscosity of varying solution concentrations of PLA.

a) Authors to whom correspondence should be addressed. Electronic email: kkparker@seas.harvard.edu and lm@seas.harvard.edu.